\documentclass[sigconf]{acmart}
\usepackage{enumitem}
\usepackage{graphicx}
\usepackage{colortbl}
\usepackage{multirow}
\usepackage{xcolor}
\usepackage{tcolorbox} 
\usepackage{booktabs}
\usepackage{subfig}
\usepackage{lipsum}
\usepackage{algorithm}
\usepackage{algorithmic}
\usepackage{tabularx}

\AtBeginDocument{%
  }

\copyrightyear{2025}
\acmYear{2025}
\setcopyright{acmlicensed}\acmConference[SIGIR '25]{Proceedings of the 48th International ACM SIGIR Conference on Research and Development in Information Retrieval}{July 13--18, 2025}{Padua, Italy}
\acmBooktitle{Proceedings of the 48th International ACM SIGIR Conference on Research and Development in Information Retrieval (SIGIR '25), July 13--18, 2025, Padua, Italy}
\acmDOI{10.1145/3726302.3729893}
\acmISBN{979-8-4007-1592-1/2025/07}
\begin{document}

\title{Agentic Feedback Loop Modeling Improves Recommendation and User Simulation}

\settopmatter{authorsperrow=4}

\author{Shihao Cai}
\orcid{0009-0009-6894-364X}
\affiliation{%
  \institution{University of Science and Technology of China}
  \city{Hefei}
  \state{Anhui}
  \country{China}
}
\email{caishihao@mail.ustc.edu.cn}

\author{Jizhi Zhang}
\orcid{0000-0002-0251-465X}
\affiliation{%
  \institution{University of Science and Technology of China}
  \city{Hefei}
  \state{Anhui}
  \country{China}
}
\email{cdzhangjizhi@mail.ustc.edu.cn}

\author{Keqin Bao}
\orcid{0009-0001-5910-0204}
\affiliation{%
  \institution{University of Science and Technology of China}
  \city{Hefei}
  \state{Anhui}
  \country{China}
}
\email{baokq@mail.ustc.edu.cn}

\author{Chongming Gao}
\authornote{Corresponding authors.}
\orcid{0000-0002-5187-9196}
\affiliation{%
  \institution{University of Science and Technology of China}
  \city{Hefei}
  \state{Anhui}
  \country{China}
}
\email{chongming.gao@gmail.com}

\author{Qifan Wang}
\orcid{0000-0002-7570-5756}
\affiliation{%
  \institution{Meta AI}
  \city{Menlo Park}
  \state{CA}
  \country{United States}
}
\email{wqfcr@fb.com}

\author{Fuli Feng}
\orcid{0000-0002-5828-9842}
\affiliation{%
  \department{MoE Key Lab of BIPC,}
  \institution{University of Science and Technology of China}
  \city{Hefei}
  \state{Anhui}
  \country{China}
}
\email{fulifeng93@gmail.com}

\author{Xiangnan He}
\authornotemark[1]
\orcid{0000-0001-8472-7992}
\affiliation{%
  \department{MoE Key Lab of BIPC,}
  \institution{University of Science and Technology of China}
  \city{Hefei}
  \state{Anhui}
  \country{China}
}
\email{xiangnanhe@gmail.com}

\renewcommand{\shortauthors}{Shihao Cai et al.}

\begin{abstract}
Large language model-based agents are increasingly applied in the recommendation field due to their extensive knowledge and strong planning capabilities.
While prior research has primarily focused on enhancing either the recommendation agent or the user agent individually, the collaborative interaction between the two has often been overlooked.
 Towards this research gap, we propose a novel framework that emphasizes the feedback loop process to facilitate  the collaboration between the recommendation agent and the user agent.
Specifically, the recommendation agent refines its understanding of user preferences by analyzing the feedback from the user agent on the item recommendation.
Conversely, the user agent further identifies potential user interests based on the items and recommendation reasons provided by the recommendation agent.
This iterative process enhances the ability of both agents to infer user behaviors, enabling more effective item recommendations and more accurate user simulations.
Extensive experiments on three datasets demonstrate the effectiveness of the agentic feedback loop: the agentic feedback loop yields an average improvement of 11.52\% over the single recommendation agent and 21.12\% over the single user agent.
Furthermore, the results show that the agentic feedback loop does not exacerbate popularity or position bias, which are typically amplified by the real-world feedback loop, highlighting its robustness.
The source code is available at \url{https://github.com/Lanyu0303/AFL}.
\end{abstract}




\maketitle
\begin{figure}[!t]
\centering
\includegraphics[width=0.9\columnwidth]{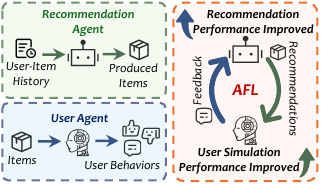}

\caption{
The comparison between the recommendation agent, the user agent, and the AFL.
The recommendation agent typically recommends items based on user-item history, whereas the user agent generally simulates user behavior towards these items.
AFL concurrently develops both a recommendation agent and a user agent,  emphasizing the interaction and collaboration between two agents.
}
\label{fig:comp}
\end{figure}
\section{Introduction}

In recent years, substantial efforts have been dedicated to developing agents based on large language models (LLMs), aiming at simulating human behavior to enhance the performance across a wide range of tasks~\cite{llm_agent_survey_chen}. These LLM-based agents typically integrate memory modules~\cite{llm_agent_memory_survey}, utilize tools~\cite{easytool}, and perform advanced reasoning~\cite{llm_agent_survey_chen, llm_planer}. These capabilities enable them to store and leverage memories for decision-making, retrieve additional information through tool usage, and apply logical reasoning to tackle complex tasks more effectively.
Building on these advantages, researchers have recently begun exploring the application of LLM-based agents in the recommendation domain~\cite{recmind,agent4rec,agentcf}.

Current applications of LLM-based agents in the recommendation field can be generally summarized into two categories:

\begin{itemize}[leftmargin=*]
 \item As to the recommendation task, the LLM-based recommendation agent~\cite{recmind,macrec} draws on the world knowledge embedded in LLMs and improves the performance through advanced capabilities such as tool use and logical reasoning~\cite{recmind,macrec,DBLP:conf/sigir/Shi0ZGLZWF24}. 

\item As to the user simulation task, the LLM-based user agent~\cite{agent4rec,recagent} leverages the human behavior modeling capabilities of LLMs to simulate user actions in recommendation systems, such as liking, disliking, and commenting. Consequently, the user agent can be used to evaluate the performance of the recommendation system, infer user interests, and generate user data for training recommendation models~\cite{agent4rec, agentcf, recllm}.
\end{itemize}

Existing research primarily focuses on optimizing either the recommendation agent or the user agent separately, overlooking the critical role of the feedback loop between the user and the recommender.
However, in real-world recommendation scenarios, the recommender aids users in discovering their interests and preferences. Simultaneously, users, through multi-round interactions with the recommender, provide feedback that enables the system to better understand their preferences. This reciprocal influence between the user and the recommender forms the feedback loop within a recommendation system~\cite{ELIXIR}.
The interactive and reciprocal nature of this feedback loop aligns well with the strengths of LLM-based agents, which excel in interaction and memory capabilities. This synergy motivates us to explore integrating the feedback loop into the optimization of both recommendation and user agents, aiming to enhance them simultaneously.

To this end, we introduce the \textbf{A}gentic \textbf{F}eedback \textbf{L}oop (AFL) modeling.
AFL simultaneously constructs both a recommendation agent and a user agent, using textual communication to simulate the feedback loop.
In each iteration, the recommendation agent suggests an item and provides a rationale. The user agent then responds with feedback, indicating whether it likes the item. If the user agent is satisfied, the loop terminates; otherwise, both the recommendation rationale and user feedback are stored in memory, and the process repeats.
AFL relies exclusively on memory to record interactions and update both agents, making it simple and broadly applicable. It is not limited to any specific agents and can be integrated with almost any recommendation or user agent equipped with memory.
By leveraging interaction history stored in memory, the recommendation agent can identify the shortcomings of previous suggestions and better infer user preferences, leading to improved recommendations.
Similarly, the user agent can discover potential user interests from the interaction history, allowing it to adjust its user modeling accordingly.

To assess the effectiveness of the proposed framework, we conducted comprehensive experiments on three widely used recommendation datasets: LastFM~\cite{lastfm}, Steam~\cite{sasrec}, and MovieLens~\cite{movielens}.
The experimental results show that AFL simultaneously enhances the performance of both the recommendation agent and the user agent, with larger improvements as the maximum number of iterations increases. 
Furthermore, unlike real-world feedback loops that often amplify popularity and position biases~\cite{DBLP:conf/cikm/MansouryAPMB20}, the experimental results show that AFL does not exacerbate these biases, demonstrating its robustness.

In conclusion, our main contributions are summarized as follows:
\begin{itemize}[topsep=0pt,itemsep=0pt,parsep=0pt,leftmargin=*]
    \item To our knowledge, our work is the first to highlight the significance of modeling the feedback loop between the recommendation agent and the user agent in LLM-based recommendation.
    \item We propose a novel framework, AFL, which establishes an agentic feedback loop to facilitate the cooperation and reciprocity between the recommendation agent and the user agent.
    \item Extensive experiments validate the effectiveness of the proposed AFL approach, underscoring the potential and importance of the feedback loop in agent-based recommendation systems.
    
\end{itemize}

\section{Related Work}
In this section, we delve into related studies from two perspectives: LLMs for recommendation and agent-based recommendation.

\subsection{LLMs for Recommendation}
Due to LLMs' extensive knowledge and strong reasoning capabilities, researchers have begun exploring their application in recommendation systems~\cite{llm_rec_survey_chen, tutorial1, personalized_survey, lin2024bridging}.
Previous work can be roughly categorized into using LLMs for feature enhancement and using LLMs for direct recommendation~\cite{rah}.

For feature enhancement, some researchers use LLMs to augment raw data, thereby supplying additional information for recommendation models~\cite{know_aug}.
What's more, several studies propose to augment traditional models with LLM tokens or embeddings~\cite{DBLP:conf/kdd/HouMZLDW22}, which can leverage the world knowledge of LLMs to assist traditional recommendation models in learning the underlying connections between items and users.

For direct recommendation, some work directly utilized LLMs' strong capabilities such as reasoning recommendation based on user information and interacted items~\cite{he2023large}, to further explore the potential of LLMs in recommendation.
Further, some studies propose fine-tuning LLMs on recommendation data, to acquire recommendation-specific knowledge and enhance recommendation performance~\cite{tallrec, zhang2023collm, chen2024softmax, gao2024sprec, gao2025flower}.
However, conducting such fine-tuning can be costly and may potentially hurt the unique capabilities of LLMs, such as reasoning and planning~\cite{bao2023bi, tutorial1}.
Thus, we focus on using LLM-based agents for recommendation in this paper, since they can acquire recommendation knowledge and maintain strong specific abilities like reasoning for better recommendation.

\subsection{Agent-based Recommendation} 
Leveraging the extensive capabilities of LLMs, agents possess powerful capabilities such as planning and execution, allowing them to tackle a wide range of complex tasks~\cite{llm_agent_survey_chen}.
Thus, many studies have explored the application of LLM-based agents in recommendation systems, which can be roughly divided into two categories: for recommendation and for user simulation.

The recommendation agent focuses on tackling recommendation tasks~\cite{recmind,macrec}.
For example, RecMind~\cite{recmind} enhances performance by employing self-inspiring and tool-calling mechanisms, and is capable of addressing various recommendation tasks, including sequential recommendation and rating prediction.
What's more, MACRec~\cite{macrec} introduces multi-agent collaboration, where different agents are assigned specific roles through role-playing, such as user analyst, searcher and item analyst.
RAH~\cite{rah} positions the agent as an intermediary assistant between multiple traditional recommendation systems and users, building a user-centred framework.

The user agent focuses on simulating user behavior~\cite{agent4rec,agentcf}.
Agent4Rec~\cite{agent4rec} constructs a user agent with a profile module, memory module, and action module, capable of simulating user behavior such as viewing, rating, and providing feedback, which can assist in testing recommendation systems.
RecLLM~\cite{recllm} builds a controllable LLM-based user simulator that can be integrated into the conversational recommender system to generate synthetic conversation data.
Furthermore, RecAgent~\cite{recagent} introduces multiple user agents, which are capable of simulating multiple users' chatting and broadcasting behavior.
What's more, AgentCF~\cite{agentcf} constructs user agents and item agents to simulate user-item interactions.
However, the existing recommendation agent and user agent approaches have not considered the feedback loop between them, which is a crucial feature in recommendation systems, to simultaneously enhance the performance for both the recommendation and user simulation.

\section{Method}
In this section, we present the proposed agentic feedback loop framework named AFL.
The overview framework of our proposed AFL is depicted in Figure~\ref{fig:pipeline}.

\subsection{Overview}
\label{sec:overview}
\begin{figure}[t]
\centering
\includegraphics[width=0.8\columnwidth]{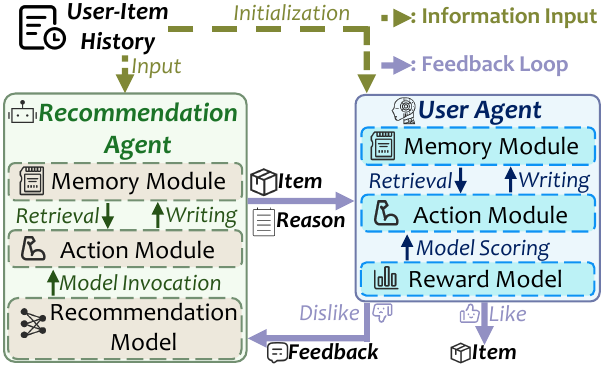}
\caption{Overview of the proposed AFL method. 
The user-item history initializes the user agent and serves as input for the recommendation agent. The recommendation agent then suggests an item with a reason. If the user agent approves, the process ends. Otherwise, the user agent provides feedback to help refine future recommendations.}
\label{fig:pipeline}
\end{figure}

As shown in Figure~\ref{fig:pipeline}, AFL consists of: 
(1) a recommendation agent (Section~\S\ref{sec:rec_agent}), 
(2) a user agent (Section~\S\ref{sec:user_agent}), and 
(3) the feedback loop between the two agents (Section~\S\ref{sec:feedback_loop}).

The input to AFL is the user-item interaction history, formalized as $[I_1, I_2, \cdots, I_n]$, where $I_i$ represents the $i$-th item the user interacts with. The output is AFL's prediction of the next item, $I_{n+1}$, that the user is likely to interact with. 
Specifically, the user-item interaction history initializes the user agent and serves as input to the recommendation agent. Based on this interaction history, the recommendation agent generates a recommended item along with the corresponding rationale. The user agent then evaluates whether it likes the recommended item. If the item is deemed favorable, the user agent directly outputs the item, concluding the process. Otherwise, the process enters an iterative feedback loop: the user agent provides feedback, prompting the recommendation agent to refine its suggestions and propose new items.

\subsection{Recommendation Agent}
\label{sec:rec_agent}
The recommendation agent is powered by GPT-4o-mini~\cite{gpt4omini} and is equipped with a memory module and a recommendation module.
The memory module can store the communication history between the recommendation and user agents, denoted as $M_r$.
In practice, the communication history is formatted into a string following the template provided in Table~\ref{tab:rec_prompt_sum}, which is then utilized as the input prompt for the agent.
The recommendation module is a flexible and interchangeable recommendation model that can recommend an item, denoted as $I_m$, based on the user-item interaction history $[I_1, I_2, \cdots, I_n]$.
Its interchangeable nature allows the recommendation agent to be flexibly integrated into recommendation systems with different architectures.
It is worth noting that the recommendation model is trained in the training set, which can provide the recommendation agent with dataset-related recommendation knowledge.
In addition, to enhance the performance of the recommendation agent, we have also adopted role-playing~\cite{roleplay} and chain-of-thought~\cite{cot} approaches, which can help the agent think step by step and provide reasons for its recommendations.

In the recommendation phase, the recommendation agent will simultaneously consider memory, user-item interaction history, and the item recommended by the recommendation model to provide a new recommended item and the corresponding reason. 
This can be formally represented as $f_r(M_r, [I_1, I_2, \cdots, I_n], I_m) = (I_r, R_r)$, where $f_r$ represents the recommendation agent, $I_r$ represents the item recommended by the recommendation agent, and $R_r$ represents the reason provided by the recommendation agent.
We present a prompt template for the recommendation agent in Table~\ref{tab:rec_prompt_sum}.
\definecolor{lightblue}{RGB}{255, 208, 199}
\definecolor{lightgray}{RGB}{245, 245, 245}

\begin{table}[h!]
    \centering
    \renewcommand{\arraystretch}{1} 
    \caption{Memory template and prompt template in the Lastfm dataset for the recommendation agent.}
    \begin{tabularx}{\linewidth}{|>{\columncolor{lightblue}}X|}
        \hline
        \textbf{Memory Template} \\ 
        \hline
        \rowcolor{lightgray}
        In round \{\}, the music artist you recommended is \{\}.

The reason you gave for the recommendation is: \{\}.

The reason the user provided for not considering this to be the best recommendation is: \{\}. \\ 
        \hline
        \textbf{Prompt Template} \\ 
        \hline
        \rowcolor{lightgray}
        You are a music artist \textbf{recommendation system}.

Refine the user's listening history to \textbf{predict the most likely music artist} he/she will listen to next from the candidate list.

Here is the \textbf{history of communication} between you and the user: \{\}.

\textbf{Another recommendation model} has suggested a music artist for your reference: \{\}.

\textbf{Some useful tips: }

1. You need to first \textbf{give the reasons}, and then provide the recommended music artist.

2. The recommended music artist must be on the candidate list.

You must follow this output format: 

\textbf{Reason}: <your reason example>

\textbf{Item}: <item example> \\ 
        \hline
    \end{tabularx}
    
    \label{tab:rec_prompt_sum}
\end{table}

\subsection{User Agent}
\label{sec:user_agent}
The user agent has a memory module $M_u$ similar to the recommendation agent's and also employs role-playing and chain-of-thought approaches.
The memory template of the user agent is presented in Table~\ref{tab:user_prompt_sum}.
Unlike the recommendation agent, the user agent uses a reward model to assign a numerical score to each recommended item, indicating its relevance to the user's interaction history.
It is important to emphasize that the reward model remains fixed, as its primary goal is to achieve optimal user simulation performance, without requiring the flexibility needed by the recommendation agent to adapt to various recommendation systems.
In our experiments, we employed SASRec~\cite{sasrec} as the reward model

Through training on the training set, the reward model can generate a score that predicts the user's preference for a given item based on the user-item interaction history.
Then, the user agent will consider the reasons for the recommendation, memory, user-item interaction history, and the score from the reward model to determine whether it likes the recommended item and provide reasons.
This can be formally represented as $f_u(M_u, [I_1, I_2, \cdots, I_n], I_r, R_r, S) = (D_u, R_u)$, where $f_u$ represents the user agent, $S$ represents the score given by the reward model, $D_u$ represents the decision of the user agent and $R_u$ represents the reason provided by the user agent.
We present a prompt template for the user agent in Table~\ref{tab:user_prompt_sum}.
\definecolor{lightblue}{RGB}{255, 208, 199}
\definecolor{lightgray}{RGB}{245, 245, 245}

\begin{table}[h!]
    \centering
    \renewcommand{\arraystretch}{1} 
    \caption{Memory template and prompt template in the Lastfm dataset for the user agent.}
    \begin{tabularx}{\linewidth}{|>{\columncolor{lightblue}}X|}
        \hline
        \textbf{Memory Template} \\ 
        \hline
        \rowcolor{lightgray}
        In round \{\}, the recommended music artist is \{\}.

The reason given by the recommendation system is: \{\}

The reason you provided for not considering this the best recommendation is \{\} \\ 
        \hline
        \textbf{Prompt Template} \\ 
        \hline
        \rowcolor{lightgray}
        As a \textbf{music listener}, you've listened to the following music artists: \{\}.

Now, a recommendation system has recommended a music artist to you from a list of music artist candidates, and has provided the reason for the recommendation.

\textbf{Determine if this recommended music artist is the most preferred option} from the list of candidates based on your personal tastes and previous listening records.

Here is the \textbf{history of communication} between you and the recommendation system: \{\}

What's more, a \textbf{reward model} scores the music artist based on its relevance to your historical listening records: \{\}

\textbf{Some useful tips:}

1. You need to first \textbf{give the reasons}, and then decide whether or not the recommended music artist is the most preferred one on the candidate list for you.

2. \textbf{Summarize your own interests} based on your historical listening records to make a judgment.

3. You can \textbf{refer to the score} given by the reward model.

You must follow this output format: 

\textbf{Reason}: <your reason example>

\textbf{Decision}: <yes or no> \\ 
        \hline
    \end{tabularx}
    
    \label{tab:user_prompt_sum}
\end{table}

\subsection{Feedback Loop}
\label{sec:feedback_loop}
The pseudocode for the AFL method is provided in Algorithm~\ref{alg:uafrs}, where ``$Rec\_Model$'' represents the recommendation model of the recommendation agent and ``$Max\_Epoch$'' represents the maximum number of iterations for the feedback loop.

As described in Section~\S\ref{sec:overview}, the process begins with the recommendation agent suggesting an item to the user agent. If the user agent finds the suggested item appealing, the process terminates. Otherwise, the process enters an iterative feedback loop, during which the agents undergo further optimization. This iterative refinement leverages the user agent's feedback to improve future interactions and enhance the overall recommendation quality.

During the iterative loop phase, both the recommendation agent and the user agent begin by storing key information in memory, including the recommended item, the reasons for the recommendation, and the user agent's reasons for rejecting the item. 
This initial step ensures that both agents have a comprehensive record of the interaction for future reference.
Next, drawing upon stored memory, the recommendation agent can reanalyze and summarize the user's interests and preferences to optimize its behavior.
In particular, the recommendation agent can attempt to persuade the user agent to accept the previously recommended item or suggest another item.
On the other hand, the user agent can analyze and extract latent interests from the items and reasons provided by the recommendation agent, enhancing its ability to simulate the user's preferences and behavior. 
Subsequently, the better user agent can provide feedback on newly recommended items that better align with the user's interests, thereby helping the recommendation agent to further improve.

As a result, during the feedback loop, both the recommendation agent and the user agent undergo iterative updates. 
This iterative refinement process leads to an improvement in the reasoning abilities of the recommendation agent and the user agent, enabling them to better recommend items and more accurately simulate user behavior, respectively.
\begin{algorithm}
\caption{AFL}
\label{alg:uafrs}
\begin{algorithmic}[1]
\STATE INPUT:($[I_1, I_2, \cdots, I_n], f_r, M_r, Rec\_Model, f_u, M_u$, $Reward\_Model, Max\_Epoch$)
\STATE $E \leftarrow 1$
\STATE $I_m \leftarrow Rec\_Model([I_1, I_2, \cdots, I_n])$ \hfill $\triangleright$ Utilize the Rec\_Model
\WHILE{$E \leq Max\_Epoch$}
    \STATE $(I_r, R_r) \leftarrow f_r(M_r, [I_1, I_2, \cdots, I_n], I_m)$
    \STATE $S \leftarrow Reward\_Model(I_r)$ \hfill $\triangleright$ Utilize the Reward\_Model
    \STATE $(D_u, R_u) \leftarrow f_u(M_u, [I_1, I_2, \cdots, I_n], I_r, R_r, S)$
    \IF{$D_u$ is False}
        \STATE $M_u \leftarrow M_u \cup \{I_r, R_r, R_u\}$ \hfill $\triangleright$ Update the memory
        \STATE $M_r \leftarrow M_r \cup \{I_r, R_r, R_u\}$
        \STATE $E \leftarrow E + 1$
    \ELSE
        \STATE \textbf{break}
    \ENDIF
\ENDWHILE
\RETURN $I_r$
\end{algorithmic}
\end{algorithm}
\section{Experiments}
In this section, we conduct experiments to answer the following research questions (RQ):

\begin{itemize}[leftmargin=*]
\item \textbf{RQ1}: Can AFL enhance performance in both the recommendation task and the user behavior simulation task?
\item \textbf{RQ2}: What are the effects of the key components of AFL?
\item \textbf{RQ3}: Does AFL amplify biases in the feedback loop?
\end{itemize}

\subsection{Experimental Setup}

\subsubsection{Datasets}
We choose three widely used recommendation datasets - Lastfm~\cite{lastfm}, Steam~\cite{sasrec}, and MovieLens~\cite{movielens} for our experiments to verify the performance of the recommendation agent and the user agent.
We sort the sequences of each dataset according to time and then divide the data into training, validation, and test sets in the ratio of 8:1:1, which ensures that subsequent interactions are excluded from the training data~\cite{DBLP:journals/tois/JiS0L23}.
In practice, the training set and validation set are used for the training and validation of the recommendation agent's recommendation model and the user agent's reward model.
Notably, due to the large size of the Steam test set, we randomly sampled 200 data points to align with the number of test samples in the Lastfm and MovieLens datasets.
The statistics of these datasets are provided in Table~\ref{tab:dataset_statistics}, and the detailed information on these datasets is as follows:
\begin{itemize}[leftmargin=*]
\item \textit{Lastfm} contains a rich set of user-artist listening records collected from the Last.fm platform.
\item \textit{Steam} is a dataset of user reviews from the Steam store. Following LLaRA~\cite{llara}, we filtered out users with fewer than 20 reviews and randomly sampled one-third of the remaining users and games to ensure a manageable dataset size.
\item \textit{MovieLens} is a popular dataset for movie recommendation, containing user ratings and multiple subset sizes. To minimize API call costs, we used the MovieLens100k subset in our experiments.
\end{itemize}

\subsubsection{Evaluation Setting}
\label{sec:eval}
Since AFL builds both a recommendation agent and a user agent, our experiments evaluate the performance of both recommendation and user simulation.

\textbf{For the recommendation task}, we adopt the experimental setup of LLaRA~\cite{llara}.
Given a user's interaction history, we first combine the next item the user will interact with and 19 randomly sampled items that the user has not interacted with, forming a candidate list of 20 items.
The model's ability to identify the correct item is evaluated using HitRatio@1.

\textbf{For the user simulation task}, we refer to the experimental setup of Agent4Rec~\cite{agent4rec}.
 Each user agent is randomly assigned 20 items.
Among these, the ratio between positive items and negative items is set as $1:k$, with $k \in \{1,3,9\}$.
Here, positive items refer to those that the user has interacted with but were not used for agent initialization, while negative items correspond to those the user has not interacted with.
For each item, the user agent needs to determine whether it likes the item.
Under this setting, the response of the user agent to each item can be regarded as binary discrimination.
Thus, we can use precision, recall, and F1 scores to evaluate the performance of user simulation.
\begin{table}[t]
\centering
\caption{\textbf{Statistics of datasets.}}
\label{tab:dataset_statistics}
\begin{tabular}{lrrr}
\toprule
Dataset& \#Sequence &\#Item &\#Interaction             \\ 
\toprule
Lastfm&1,220&4,606&73,510 \\
Steam&11,938&3,581&274,726 \\
MovieLens&943&1,682&100,000 \\
\bottomrule
\end{tabular}
\end{table}
\begin{table*}[htbp]
\caption{The recommendation performance of AFL compared with ``Base Model'' and ``Rec Agent''. 
Bold indicates the best performance.
The maximum number of feedback loop iterations for AFL is 4.}
\label{tab:rec_exp}
\centering
 \renewcommand\tabcolsep{2.4pt} 
\begin{tabular}{lc|ccc|ccc|ccc}
\toprule
\multirow{2}{*}{Type} &\multirow{2}{*}{Model} & \multicolumn{3}{c|}{Lastfm} & \multicolumn{3}{c|}{Steam} & \multicolumn{3}{c}{MovieLens}\\
 & & Base Model & Rec Agent & AFL & Base Model & Rec Agent & AFL & Base Model & Rec Agent & AFL\\
\toprule
\multirow{3}{*}{Traditional}&SASRec    & 0.2869& 0.3197& \textbf{0.3770}  & 0.3800& 0.3900& \textbf{0.4100} & 0.4105& 0.4105& \textbf{0.4316} \\
                            &GRU4Rec   & 0.2787& 0.3114& \textbf{0.3770}  & 0.3750& 0.3850& \textbf{0.4100} & 0.4526& 0.4526& \textbf{0.4632} \\
                            &Caser     & 0.2705& 0.2705& \textbf{0.3443}  & 0.4200& 0.4150& \textbf{0.4500} & 0.3789& 0.3895& \textbf{0.4000} \\
\midrule
\multirow{4}{*}{LLM-based}  &MoRec     & 0.1639& 0.2131& \textbf{0.3115}  & 0.4100& 0.4200& \textbf{0.4250} & 0.3158& 0.3158& \textbf{0.3474} \\
                            &Llama3-8B & 0.2131& 0.2541& \textbf{0.2869}  & 0.1800& 0.2250& \textbf{0.3000} & 0.1368& 0.1368& \textbf{0.1684} \\
                            &GPT-4o-mini& 0.3607& 0.3607& \textbf{0.3770}  & 0.3350& 0.3400& \textbf{0.3500} & 0.1368& 0.1368& \textbf{0.1579} \\
                            &LLaRA     & 0.4426 & 0.4426& \textbf{0.4836}  & 0.4650& 0.4650& \textbf{0.4750} & 0.4842& 0.4842& \textbf{0.4947} \\
\bottomrule
\end{tabular}
\end{table*}
\subsubsection{Base Models}
\label{sec:base_model}
We have chosen a variety of models as the base models to serve as the recommendation model for the recommendation agent.
These models can be broadly categorized into traditional recommendation models and LLM-based models.

The traditional recommendation models are as follows:
\begin{itemize}[leftmargin=*]
\item \textit{SASRec}~\cite{sasrec} is an attention-based sequential model that effectively captures long-term semantic dependencies in both sparse and dense datasets.
\item  \textit{GRU4Rec}~\cite{gru4rec} is an RNN-based model that is relatively simple yet highly efficient.
\item  \textit{Caser}~\cite{caser} treats the user's historical behavior sequence as an ``image'' and utilizes CNN to extract features from this sequence.
\end{itemize}
The LLM-based models are as follows:
\begin{itemize}[leftmargin=*]
\item \textit{MoRec}~\cite{morec} improves traditional recommendation models by incorporating modality features of items.
\item  \textit{Llama3}~\cite{llama3} is one of the most popular open-source LLMs.
\item  \textit{GPT-4o-mini}~\cite{gpt4omini} is one of the most powerful commercial models, capable of handling a wide range of complex tasks.
\item  \textit{LLaRA}~\cite{llara} utilizes the hybrid item representation to combine LLMs with traditional recommendation models, and it applies a curriculum learning approach to gradually increase the complexity of training.
\end{itemize}

\subsubsection{Implementation Details}
To build agents, we utilized the ``GPT-4o-mini-2024-07-18''\footnote{\url{https://platform.openai.com/docs/models/gpt-4o-mini}.} API provided by OpenAI.
As discussed in Sections~\S\ref{sec:rec_agent} and \S\ref{sec:user_agent}, the recommendation model within the recommendation agent is flexible and adaptable, whereas the reward model of the user agent remains fixed.
In practice, the recommendation model differs among the base models outlined in Section~\S\ref{sec:base_model}, while the reward model is consistently implemented using SASRec.
For traditional models, we strictly follow~\cite{frame_seq_rec}, with a learning rate of 0.001, an embedding dimension of 64, and a batch size of 256.
What's more, we also perform a grid search over the values $[1e-3, 1e-4, 1e-5, 1e-6, 1e-7]$ to determine the optimal coefficient for L2 regularization.
For LLMs-based models, we follow LLaRA~\cite{llara} and train the models for up to 5 epochs with a batch size of 128.
Moreover, we use a warm-up strategy, starting the learning rate at $\frac{1}{100}$ of the maximum and gradually increasing it with a cosine scheduler during training.
\subsection{Effectiveness of AFL (RQ1)}
In this section, we validate the effectiveness of AFL.
Specifically, this section includes: (1) the improvement in recommendation performance (Section~\S\ref{sec:rec_exp}), (2) the improvement in user simulation performance (Section~\S\ref{sec:user_exp}), and (3) a case study illustrating the agent collaboration mechanism behind the effectiveness of the feedback loop (Section~\S\ref{sec:case_study}).

\subsubsection{Recommendation Performance}
\label{sec:rec_exp}
In this section, we explore whether AFL can improve the performance of recommendation agents equipped with various base models.
Thus, the user agent's reward model is consistently implemented using the SASRec model, while the recommendation agent’s recommendation model is varied among different base models described in Section~\S\ref{sec:base_model}.
For convenience, we define the term ``Base Model'' as directly recommending items using the original base model.
In contrast, the term ``Rec Agent'' refers to directly using the recommendation agent equipped with the corresponding base model to recommend items.
The comprehensive experimental results can be found in Table~\ref{tab:rec_exp}.

Based on the experimental results presented in Table~\ref{tab:rec_exp}, we have drawn the following findings and conclusions:
(1) First, the recommendation agent equipped with the recommendation model outperforms or performs equally well compared to the original base model in most cases. 
This finding underscores the ability of the LLM-based recommendation agent to leverage the extensive world knowledge and sophisticated reasoning capabilities inherent in LLMs to improve recommendation accuracy.
(2) Moreover, the AFL, which is based on the feedback loop, demonstrates a significantly greater improvement than the use of a standalone recommendation agent. 
This result emphasizes the critical role of the user agent's feedback in refining and optimizing the recommendation process, further improving the recommendation performance of the recommendation agent.
(3) Finally, it is important to highlight that by changing the recommendation model of the recommendation agent, AFL can improve the recommendation performance of various base models with unique frameworks, indicating that AFL has strong generalizability and transferability across different recommendation systems.
This underscores AFL's ability to adapt and enhance diverse approaches, making it a versatile and valuable framework in recommendation scenarios.

\subsubsection{User Simulation Performance}
\label{sec:user_exp}
\begin{table*}[htbp]
\caption{The user simulation performance of AFL compared with ``Reward Model'' and ``User Agent''.
Bold results indicate the best results.
The maximum number of feedback loop iterations for AFL is 4.}
\label{tab:user_exp}
\centering
\begin{tabular}{lc|ccc|ccc|ccc}
\toprule
\multirow{2}{*}{$1:k$} &\multirow{2}{*}{Method}& \multicolumn{3}{c|}{Lastfm}& \multicolumn{3}{c|}{Steam} & \multicolumn{3}{c}{Movielens} \\
 &  & Precision & Recall & F1 Score & Precision & Recall & F1 Score & Precision & Recall & F1 Score\\
\toprule
\multirow{3}{*}{1:1}&Reward Model&0.6667&0.0533&0.0988&0.7826&0.6800&0.7277&0.6929&0.3800&0.4908\\
                    &User Agent&0.8155&0.3467&0.4865&0.8031&\textbf{0.6933}&0.7422&0.7049&0.5133&0.5941\\
                    &AFL &\textbf{0.8504}&\textbf{0.5000}&\textbf{0.6297}&\textbf{0.8501}&0.6700&\textbf{0.7494}&\textbf{0.7065}&\textbf{0.5500}&\textbf{0.6185}\\
\midrule
\multirow{3}{*}{1:3}&Reward Model&0.4167&0.0571&0.1005&0.5791&0.7133&0.6393&0.5179&0.3077&0.3860\\
                    &User Agent&0.5910&0.3571&0.4452&0.6323&0.7067&0.6674&0.5114&\textbf{0.4800}&0.4952\\
                    &AFL &\textbf{0.7343}&\textbf{0.4286}&\textbf{0.5412}&\textbf{0.6815}&\textbf{0.7267}&\textbf{0.7034}&\textbf{0.8107}&0.4667&\textbf{0.5924}\\
\midrule
\multirow{3}{*}{1:9}&Reward Model&0.1667&0.0667&0.0952&0.3408&0.7667&0.4718&0.3397&0.2667&0.2988\\
                    &User Agent&0.2356&0.2667&0.2501&0.3682&\textbf{0.8167}&0.5076&0.2313&\textbf{0.5000}&0.3163\\
                    &AFL &\textbf{0.3705}&\textbf{0.4286}&\textbf{0.3974}&\textbf{0.4303}&\textbf{0.8167}&\textbf{0.5636}&\textbf{0.4410}&0.4333&\textbf{0.4371}\\
\bottomrule
\end{tabular}
\end{table*}
In this section, we investigate whether AFL can enhance the performance of the user agent.
The evaluation setup is detailed in Section~\S\ref{sec:eval}.
Specifically, we compare the performance of AFL against two baselines: using only the reward model (abbreviated as ``Reward Model'') and using only the user agent (abbreviated as ``User Agent'').
To ensure a fair comparison and avoid introducing additional information through the recommendation model of AFL's recommendation agent, we align the recommendation model with the reward model by using the same architecture, SASRec.
The detailed experimental results are provided in Table~\ref{tab:user_exp}.

Based on results demonstrated in Table~\ref{tab:user_exp}, we have following key observations:  
(1) Firstly, the user agent demonstrates superior performance compared to the original reward model across the majority of scenarios, excelling particularly in precision and recall.
This indicates that the LLM-based user agent can effectively leverage the reasoning capabilities of LLMs, utilizing the reward model's scoring mechanism to thoroughly evaluate how well an item aligns with the user's preferences and interests.
(2) Secondly, the AFL exhibits an even more significant performance improvement.
This demonstrates that the user agent can better understand the user's interests during the feedback loop.

However, it is worth noting that in some scenarios, such as when $1:k=1:3$, AFL's recall on the MovieLens dataset is lower than that of ``User Agent''.
By carefully examining the three metrics, we observe that ``User Agent'' exhibits lower precision and F1 score, suggesting a strong tendency to assume users identically like all recommended items, thereby achieving higher recall.
Conversely, AFL demonstrates a significantly higher F1 score, indicating a better simulation of user behavior. 
Therefore, we consider the slight decrease in recall to be acceptable.

\subsubsection{Case Study}
\label{sec:case_study}
\begin{figure*}[htbp]
\centering
\includegraphics[width=0.82\textwidth]{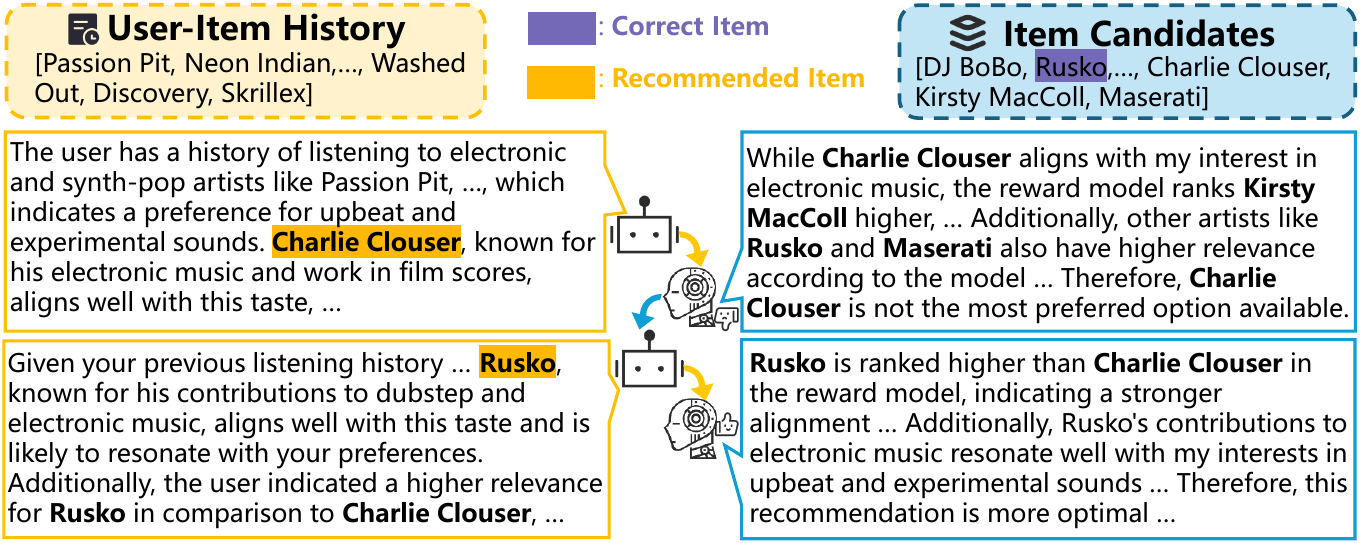}
\caption{The example of interaction between the recommendation agent and the user agent.
Given the user-item interaction history, the recommendation agent needs to identify the correct item from the list of candidate items.}
\label{fig:example}
\end{figure*}

To illustrate the effectiveness of collaboration between the recommendation agent and the user agent, we present an example in Figure~\ref{fig:example}.
In detail, the recommendation agent begins by analyzing the user's listening history, identifying a preference for upbeat and experimental sounds, and subsequently suggests the artist  ``Charlie Clouser''.
However, the user agent, utilizing insights from the reward model, determines that ``Charlie Clouser'' is not the optimal answer and expresses a stronger preference for ``Kirsty MacColl'', ``Rusko'' and ``Maserati''.
In response, the recommendation agent evaluates the reason provided by the user agent and adjusts its suggestion, recommending ``Rusko'' instead.
Finally, the user agent approved the artist recommended by the recommendation agent, and the feedback loop was terminated.

In conclusion, during the feedback loop, the recommendation agent dynamically refined its suggestions based on the user agent's feedback, ultimately delivering more accurate recommendations. 
Simultaneously, the user agent uncovered the user's interests based on the items recommended by the recommendation agent, allowing for a more precise simulation of the user's behavior.

\subsection{Impact of Key Components (RQ2)}

In this section, we validate the effectiveness of the key components of AFL.
Specifically, this section includes:
(1) the effect of feedback from the user agent (Section~\S\ref{sec:ranker_exp}),
(2) the effect of the recommendation model and the reward model (Section~\S\ref{sec:ablation_exp}),
(3) the effect of the number of the feedback loop iterations (Section~\S\ref{sec:round_exp}).

\subsubsection{Impact of User Agent Feedback}
\label{sec:ranker_exp}
Although Section~\ref{sec:rec_exp} shows the effectiveness of utilizing user agent feedback to enhance the performance of the recommendation agent, it remains unclear how this approach compares to simpler alternatives, like directly incorporating a ranker subsequent to the recommendation agent.

To address this, we conduct a comparative analysis using both SASRec and the user agent as the ranker.
In our experimental setup, the recommendation agent first generates a candidate list of five items based on the user's interaction history.
The ranker then reorders these items and selects the best one as the final output.
For clarity, the terms ``Rec Agent + SASRec'' and ``Rec Agent + User Agent'' respectively refer to the integration of SASRec and the user agent, following the recommendation agent.
The experimental results are shown in Table~\ref{tab:ranker_exp}.

Based on the experimental results in Table~\ref{tab:ranker_exp}, we have following
key observations: 
(1) Firstly, simply appending a ranker after the agent does not necessarily enhance recommendation performance.
This result is likely because the ranker assumes a dominant role in determining the final output.
Consequently, the correct items generated by the agent may be ranked lower.
(2) What's more, AFL incorporates an agentic feedback loop, where feedback from the user agent is relayed back to the recommendation agent. This mechanism enables the recommendation agent to respond to the feedback by attempting to persuade the user agent, thereby allowing errors in the user agent's judgment to be identified and corrected. Consequently, AFL achieves superior performance compared to the other three settings.
\begin{table}[t]
\caption{The HitRatio@1 of AFL compared to the method using a ranker. Bold results indicate the best results.}
\label{tab:ranker_exp}
\centering
\begin{tabular}{lccc}
\toprule
Method & Lastfm & Steam & Movielens \\
\toprule
SASRec   & 0.2869 & 0.3800 & 0.4105 \\
User Agent & 0.3197 & 0.3900 & 0.4105 \\
Rec Agent    & 0.3197 & 0.3900 & 0.4105 \\
Rec Agent + SASRec & 0.3525	 & 0.3850 & 0.3895 \\
Rec Agent + User Agent & 0.3607	 & 0.3950 & 0.4000 \\
\midrule
AFL        & \textbf{0.3770} & \textbf{0.4100} & \textbf{0.4316} \\
\bottomrule
\end{tabular}
\end{table}
\subsubsection{Impact of Recommendation Model and Reward Model}
\label{sec:ablation_exp}
To evaluate the effectiveness of the recommendation model and the reward model in the two agents of AFL, we assess the recommendation performance under various settings. 
The term ``AFL w/o Rec Model'' indicates that the recommendation agent is unable to access the recommendation model, while ``AFL w/o Reward Model'' indicates that the user agent is unable to access the reward model. 
Finally, ``AFL w/o Both'' refers to a setting where neither agent can access their respective models.
For a fair comparison, we fix both the recommendation model and the reward model as SASRec models.
The detailed experimental results can be found in Table~\ref{tab:ablation_exp}.

Based on the experimental results shown in Table~\ref{tab:ablation_exp}, we can draw several key conclusions:
(1) Firstly, the integration of the recommendation model significantly improves recommendation performance, underscoring its effectiveness in improving the recommendation agent's ability to cater to user preferences.
(2) What's more, it is worth noting that equipping the user agent with a reward model can also enhance recommendation performance.
This suggests that a better user agent enables the recommendation agent to more accurately understand user preferences within the feedback loop, thereby optimizing the overall recommendation process.
(3) Lastly, the simultaneous incorporation of both the recommendation model and the reward model results in even greater performance improvements.
This not only further reaffirms the necessity of both the recommendation model and reward model but also highlights the synergistic effect of combining an advanced recommendation agent with an advanced user agent.
\begin{table}[t]
\caption{Comparison of HitRatio@1 under different settings. Bold results indicate the best results.}
\label{tab:ablation_exp}
\centering
\begin{tabular}{lccc}
\toprule
Method & Lastfm & Steam & Movielens \\
\toprule
AFL        & \textbf{0.3770} & \textbf{0.4100} & \textbf{0.4316} \\
\midrule
AFL w/o Rec Model    & 0.3525 & 0.3950 & 0.4000 \\
AFL w/o Reward Model & 0.3689 & 0.4000 & 0.4211 \\
AFL w/o Both        & 0.3443 & 0.3250 & 0.2105 \\
\bottomrule

\end{tabular}
\end{table}

\subsubsection{Impact of Feedback Loop Iterations}
\label{sec:round_exp}
\begin{figure}[t]
    \centering
    \subfloat[]{\includegraphics[width=0.49\linewidth]{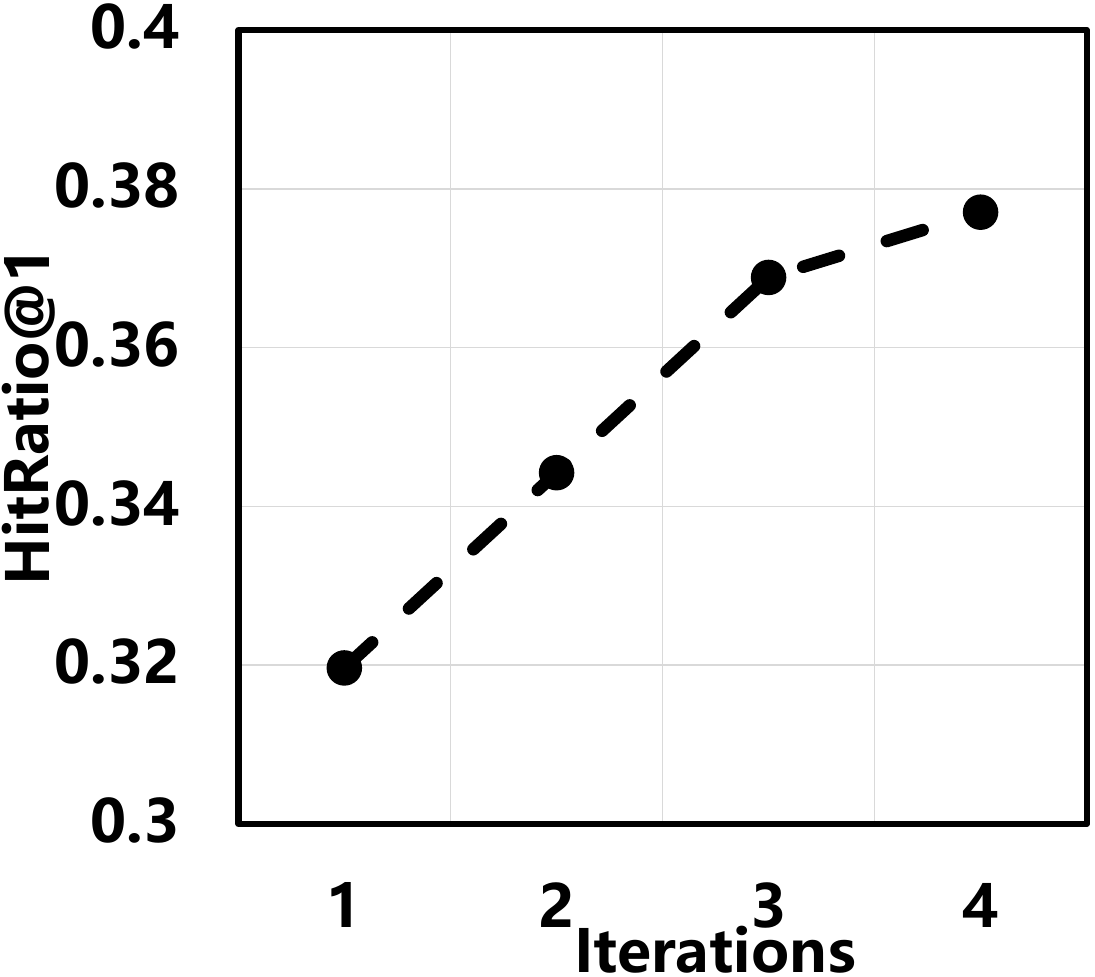}\label{fig:round_rec}}\hfill
    \subfloat[]{\includegraphics[width=0.51\linewidth]{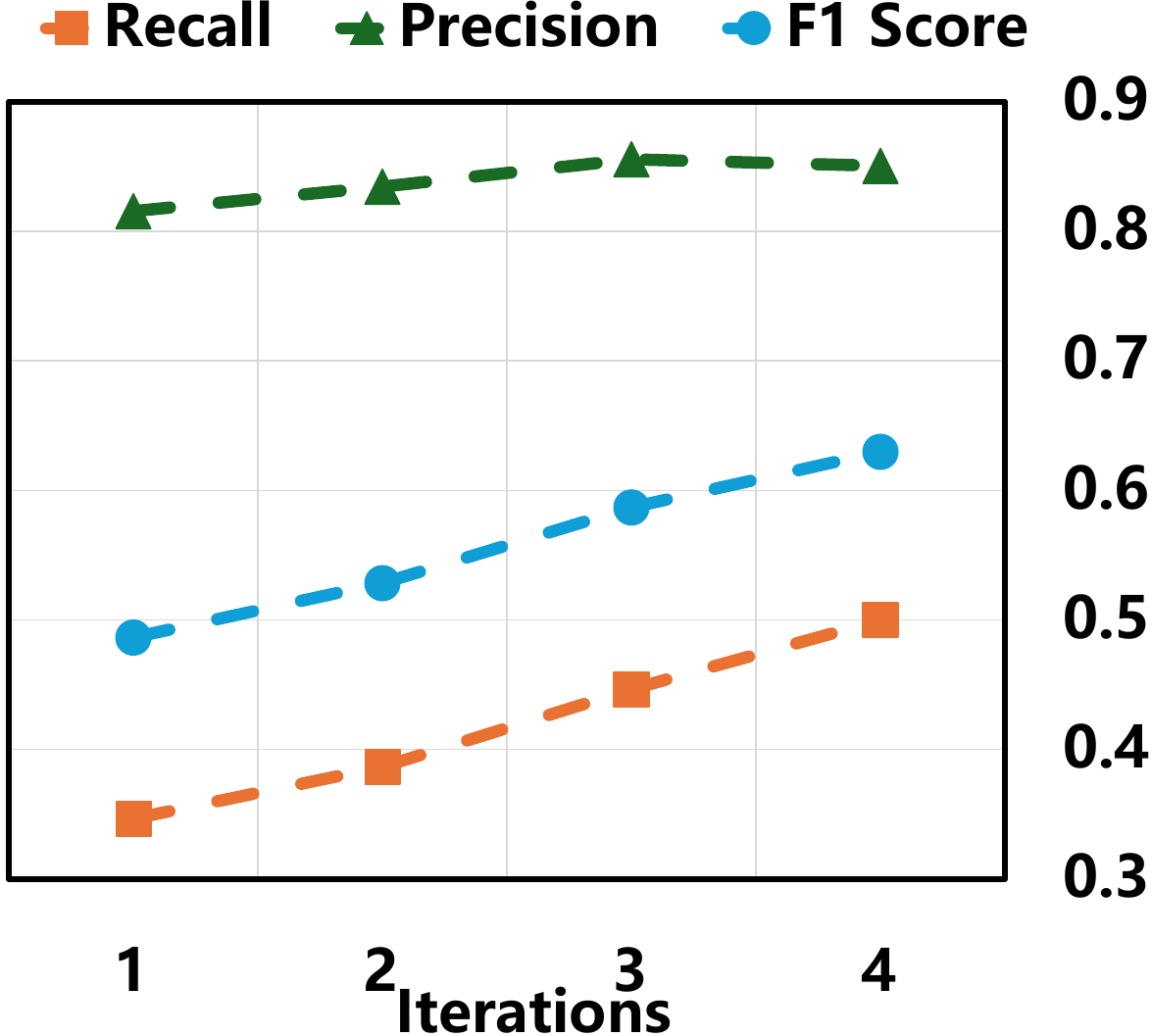}\label{fig:round_user}}
    \caption{(a) Recommendation performance with increased iterations.
(b) User simulation performance with increased iterations.
$1:k$ is set to $1:1$.
    }
    \label{fig:round}
\end{figure}

In this section, we investigate the impact of the number of feedback loop iterations.
In particular, we fix the recommendation model and the reward model as the SASRec model and evaluate the changes in the recommendation performance and user simulation performance in the Lastfm data set.
We present the curves of recommendation performance and user simulation performance as they change with increased iterations in Figure~\ref{fig:round_rec} and Figure~\ref{fig:round_user}, respectively.

\textbf{Recommendation Performance:}
The experimental results presented in Figure~\ref{fig:round_rec} reveal a clear trend: as the number of feedback loop iterations increases, the recommendation performance steadily improves.

\textbf{User Simulation Performance:}
As shown in Figure~\ref{fig:round_user}, the performance of the user simulation exhibits consistent enhancement with an increasing number of feedback loop iterations. This trend underscores the positive correlation between iterative feedback and the refinement of simulation behavior.

\textbf{Cost-Performance Tradeoff Analysis:}
It is important to note that the improvement ratio for HitRatio@1 diminishes as the number of feedback loop iterations increases, and the precision at four iterations is slightly lower than that at three iterations.
Given that increasing the number of feedback loop iterations results in higher API costs, more iterations do not necessarily lead to better outcomes.
Therefore, it is crucial to strike a balance between performance gains and API costs.
\subsection{Biases in the Feedback Loop (RQ3)}
In this section, we explore whether the agentic feedback loop amplifies two common biases found in real-world feedback loops: 
(1) popularity bias (Section~\S\ref{sec:popularity_bias}) and (2) position bias (Section~\S\ref{sec:position_bias}).
\subsubsection{Popularity Bias.}
\label{sec:popularity_bias}
As highlighted in previous research~\cite{DBLP:conf/cikm/MansouryAPMB20}, feedback loops in real-world recommendation systems are often affected by popularity bias, where the system disproportionately favors popular items at the expense of lesser-known ones \cite{gao2023alleviating,gao2023cirs}.
Thus, we investigate whether the agentic feedback loop, as simulated by the interaction between the recommendation agent and the user agent, amplifies this popularity bias.

\begin{figure}[t]
\centering
\includegraphics[width=1\columnwidth]{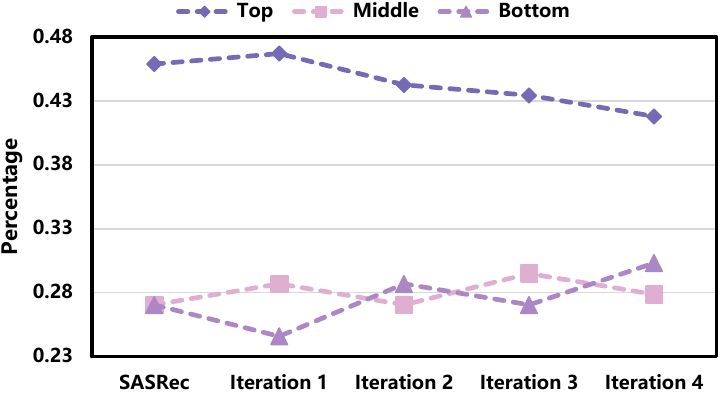}
\caption{The distribution of the three popularity categories in the Lastfm dataset under different settings.}
\label{fig:popularity}
\end{figure}

Specifically, we start by calculating the popularity of items in the Lastfm dataset by counting the number of user interactions associated with each item.
Following previous research~\cite{top20_bias}, we consider the top 20\% most popular items as ``popular items.'' 
Furthermore, we divide the remaining items into two categories: the middle 20\%–50\% and the bottom 50\% based on their popularity.
Finally, we analyze the distribution of recommended items across these three categories under varying iterations of the feedback loop.
To facilitate a better comparison, we also include the categorization of items recommended by SASRec as a baseline.
We present the results in Figure~\ref{fig:popularity}.

Based on the experimental results presented in Figure~\ref{fig:popularity}, we derive the following key observations:
(1) During the first iteration of the feedback loop, the recommendation agent generates items independently, without incorporating feedback from the user agent. As a result, its performance is primarily influenced by the underlying recommendation model (SASRec) and the inherent bias of the LLM toward popular items.
(2) From the second to fourth iterations, the feedback provided by the user agent allows the recommendation agent to better align with user preferences. This leads to recommendations that include a greater proportion of less popular items.
(3) Overall, AFL demonstrates a positive impact in mitigating popularity bias.

\subsubsection{Position Bias.}
\label{sec:position_bias}
\begin{figure}[t]
\centering
\includegraphics[width=0.95\columnwidth]{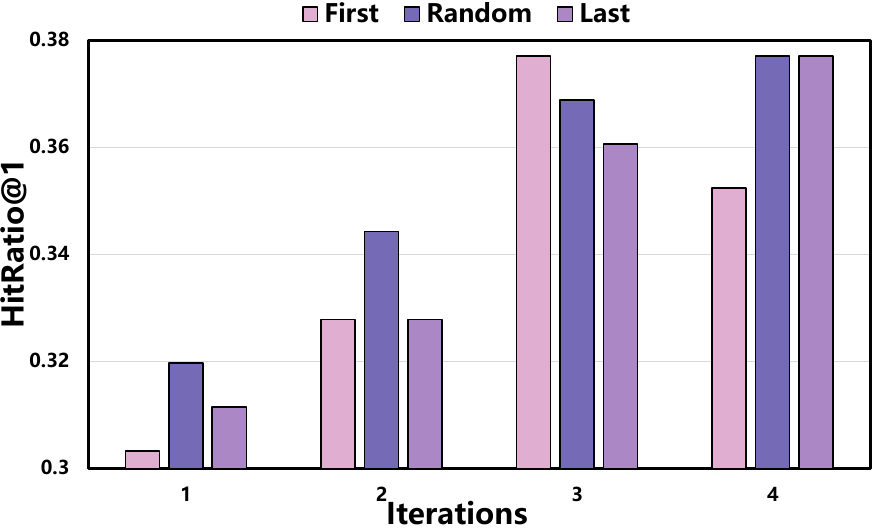}
\caption{Recommendation performance in the Lastfm dataset under three position settings.}
\label{fig:position}
\end{figure}

Notably, LLM-based recommendations can be affected by the positional arrangement of items within a sequence.
To investigate this, we design three experimental settings using the Lastfm dataset: (1) the correct item is positioned first in the list of candidate items (denoted as ``First''), (2) the correct item is placed at a random position in the list (denoted as ``Random''), and (3) the correct item is positioned last in the list (denoted as ``Last'').
If the recommendation agent relies more on positional information rather than user preferences, the ``First'' and ``Last'' settings are expected to significantly outperform the ``Random'' setting, as the agent only needs to blindly output either the first or the last item.
We present the recommendation performance for these three settings under varying maximum numbers of feedback loop iterations in Figure~\ref{fig:position}.  

Based on the experimental results shown in Figure~\ref{fig:position}, we summarize the following findings:  
(1) First, in most cases, ``Random'' outperforms both ``First'' and ``Last'', suggesting that AFL does not heavily depend on positional information.  
(2) What's more, as the number of feedback loop iterations increases, the optimal performance achievable by the three position settings becomes identical, indicating that AFL is capable of resisting positional interference. 
(3) In conclusion, AFL demonstrates resistance to location bias by learning user preferences rather than location information through the collaboration of the recommendation agent and the user agent.

\section{Conclusion and Future Work}
In this study, we propose a novel framework called AFL, which builds an agentic feedback loop to enhance both the recommendation agent and the user agent.
Our framework is simple and generic: it is not limited to the specific agents designed in this paper, but is intended to serve as a guideline for fostering collaborative and mutually beneficial interactions between the recommendation agent and the user agent.
This work highlights the collaboration between recommendation and user agents, opening up new research avenues of mutually beneficial multi-agent cooperation for recommendation and user simulation.

In the future, we will explore more efficient feedback mechanisms.
While our current approach updates agents by modifying their memory, it would be valuable to leverage feedback for training agents directly, which could help agents better understand and adapt to user interests.
Moreover, we are particularly interested in implementing an agentic feedback loop between a single recommendation agent and multiple user agents --- a setting that better reflects real-world recommendation scenarios but has received relatively limited attention in the field.
To address this gap, we need to develop effective methods for modeling multiple personalized user agents and enabling the recommendation agent to balance both the common and personalized preferences among users.
Another emerging direction is to further explore ways to reduce biases in agentic feedback loops, such as popularity bias and position bias.

\begin{acks}
This work is supported by the National Key Research and Development Program of China (2022YFB3104701), the National Natural Science Foundation of China (62402470, 62272437, U24B20180), Anhui Provincial Natural Science Foundation (2408085QF189), and the advanced computing resources provided by the Supercomputing Center of the USTC.
\end{acks}

\bibliographystyle{ACM-Reference-Format}
\balance
\bibliography{sample-base}

\end{document}